\documentclass[a4paper,12pt]{article}

\begin{document}
\makeatletter
\renewcommand{\theequation}{\thesection.\arabic{equation}}
\@addtoreset{equation}{section}
\makeatother

\title{Holographic Relation in Yang's Quantized Space-Time Algebra and Area-Entropy Relation in $D_0$ Brane Gas System}
\author{\large Sho Tanaka\footnote{Em. Professor of Kyoto University, E-mail: st-desc@kyoto.zaq.ne.jp }
\\[8 pt]
 Kurodani 33-4, Sakyo-ku, Kyoto 606-8331, Japan}
\date{}
\maketitle

\vspace{-10cm}
\rightline{}
\vspace{12cm}
\thispagestyle{empty}
 In the preceding paper, we derived a kind of kinematical holographic relation (KHR) in the Lorentz-
covariant Yang's quantized space-time algebra (YSTA). It essentially reflects the fundamental nature 
of the noncommutative geometry of YSTA and its representation, that is, a definite kinematical reduction of 
spatial degrees of freedom in comparison with the ordinary lattice space. On the basis of the relation and 
its extension to various spatial dimensions, we derive a new area-entropy relation in a simple $D_0$ brane gas system 
subject to YSTA, following the idea of M-theory. Furthermore, we make clear its inner relation with the 
Bekenstein-Hawking area-entropy relation in connection with Schwarzschild black hole.

\vspace{2\baselineskip}

\newpage

\section{\normalsize Introduction}

In the preceding paper, $^{[1]}$ referred hereafter as I, we derived a kind of holographic relation in 
the Lorentz-covariant Yang's quantized space-time algebra(YSTA),$^{[1],[2],[3]}$ which we called the kinematical 
holographic relation (KHR). As was emphasized in I, the relation essentially reflects the fundamental nature of 
the noncommutative geometry of YSTA, that is, a definite kinematical reduction of spatial degrees of freedom in 
comparison with the ordinary lattice space. As will be shown in the present paper, this relation seems also to 
give an important clue to resolve the long-pending problem encountered in the Bekenstein-Hawking area-entropy 
relation$^{[4]}$ or the holographic principle,$^{[5]}$ that is, the apparent gap between the degrees of 
freedom of any bounded spatial region associated with entropy and of its boundary area. 

In addition to the last problem, in the arguments of the holographic principle, the limit of the present local 
field theory has been discussed, as seen, for instance, in the unified regularization or cutoff of UV/IR 
divergences. With respect to this problem, as was emphasized in refs. [1], [2], YSTA which is 
intrinsically equipped with short- and long-scale parameters, $\lambda$ and $R$, gives a finite number 
of spatial degrees of freedom for any finite spatial region and provides a basis for the field theory free 
from ultraviolet- and infrared-divergences. 

In fact, we found in I, the following form of kinematical holographic relation (KHR) in YSTA,
\begin{eqnarray} 
\hspace{-3cm} [KHR] \hspace{2cm}       n^L_{\rm dof}= {\cal A} / G,
\nonumber
\end{eqnarray}
that is, the proportional relation between $n^L_{\rm dof}$ and ${\cal A}$ with proportional constant $G$, where 
$n^L_{\rm dof}$ and ${\cal A}$, respectively, denote the number of degrees of freedom of any spherical bounded 
spatial region with radius $L$ in Yang's quantized space-time and the boundary area in unit of $\lambda.$   

In this paper, we derive a new area-entropy relation [AER] on the basis of the above [KHR] and make clear 
its inner relation with the ordinary Bekenstein-Hawking area-entropy relation in connection with 
Schwarzschild black hole. It will be made through a simple $D_0$ brane gas model$^{[6]}$ on Yang's quantized 
space-time according to the idea of M-theory,$^{[7]}$ with the aid of a kind of Gedanken-experiment on the present 
static toy model. 

The present paper is organized as follows. In Sec.\ 2, we briefly recapitulate Yang's quantized space-time 
algebra (YSTA) and its representations. Sec.\ 3 is devoted to the recapitulation of the kinematical holographic relation 
(KHR) and to its extension to the lower-dimensional bounded regions, $V_d^L$. In section 4, we introduce a simple $D_0$ 
brane (D-particle) gas model on $V_d^L$ and find a new area-entropy relation in the system in connection 
with Schwarzschild black hole. In the final section, we discuss the inner relation between our area-entropy relation 
based on KHR in YSTA and the ordinary Bekenstein-Hawking area-entropy relation and point out our future task beyond 
the present simple $D_0$ brane gas model.

\section{\normalsize Yang's Quantized Space-Time Algebra (YSTA) and \break Its Representations}  

\subsection{\normalsize Yang's Quantized Space-Time Algebra (YSTA) }

Let us first recapitulate briefly the Lorentz-covariant Yang's quantized space-time algebra (YSTA).
$D$-dimensional Yang's quantized space-time algebra is introduced$^{[1],[2]}$ as the 
result of the so-called Inonu-Wigner's contraction procedure with two contraction parameters, $R$ and 
$\lambda$, from $SO(D+1,1)$ algebra with generators $\hat{\Sigma}_{MN}$; 
\begin{eqnarray}
 \hat{\Sigma}_{MN}  \equiv i (q_M \partial /{\partial{q_N}}-q_N\partial/{\partial{q_M}}),
\end{eqnarray}
which work on $(D+2)$-dimensional parameter space  $q_M$ ($M= \mu,a,b)$ satisfying  
\begin{eqnarray}
             - q_0^2 + q_1^2 + \cdots + q_{D-1}^2 + q_a^2 + q_b^2 = R^2.
\end{eqnarray}
 
Here, $q_0 =-i q_D$ and $M = a, b$ denote two extra dimensions with space-like metric signature.

$D$-dimensional space-time and momentum operators, $\hat{X}_\mu$ and $\hat{P}_\mu$, 
with $\mu =1,2,\cdots,D,$ are defined in parallel by
\begin{eqnarray}
&&\hat{X}_\mu \equiv \lambda\ \hat{\Sigma}_{\mu a}
\\
&&\hat{P}_\mu \equiv \hbar /R \ \hat{\Sigma}_{\mu b},   
\end{eqnarray}
together with $D$-dimensional angular momentum operator $\hat{M}_{\mu \nu}$
\begin{eqnarray}
   \hat{M}_{\mu \nu} \equiv \hbar \hat{\Sigma}_{\mu \nu}
\end{eqnarray} 
and the so-called reciprocity operator
\begin{eqnarray}
    \hat{N}\equiv \lambda /R\ \hat{\Sigma}_{ab}.
\end{eqnarray}
Operators  $( \hat{X}_\mu, \hat{P}_\mu, \hat{M}_{\mu \nu}, \hat{N} )$ defined above 
satisfy the so-called contracted algebra of the original $SO(D+1,1)$, or Yang's 
space-time algebra (YSTA):
\begin{eqnarray}
&&[ \hat{X}_\mu, \hat{X}_\nu ] = - i \lambda^2/\hbar \hat{M}_{\mu \nu}
\\
&&[\hat{P}_\mu,\hat{P}_\nu ] = - i\hbar / R^2\ \hat{M}_{\mu \nu}
\\
&&[\hat{X}_\mu, \hat{P}_\nu ] = - i \hbar \hat{N} \delta_{\mu \nu}
\\
&&[ \hat{N}, \hat{X}_\mu ] = - i \lambda^2 /\hbar  \hat{P}_\mu
\\
&&[ \hat{N}, \hat{P}_\mu ] =  i \hbar/ R^2\ \hat{X}_\mu,
\end{eqnarray}
with familiar relations among ${\hat M}_{\mu \nu}$'s omitted.

\subsection{Quasi-Regular Representation of YSTA}

Let us further recapitulate briefly the representation$^{[1],[2]}$ of YSTA for the subsequent 
consideration in section 4. First, it is important to notice the following elementary fact that ${\hat\Sigma}_{MN}$ 
defined in Eq.(2.1) with $M, N$ being the same metric signature have discrete eigenvalues, i.e., $0,\pm 1 ,
\pm 2,\cdots$, and those with $M, N$ being opposite metric signature have continuous eigenvalues,
$\footnote{The corresponding eigenfunctions are explicitly given in ref. [9].}$ consistently with 
covariant commutation relations of YSTA. This fact was first emphasized by Yang$^{[3]}$ in connection with 
the preceding Snyder's quantized space-time.$^{[8]}$ This conspicuous aspect is well understood by means of 
the familiar example of the three-dimensional angular momentum in quantum mechanics, where individual components, 
which are noncommutative among themselves, are able to have discrete eigenvalues, consistently with the 
three-dimensional rotation-invariance. 
 
This fact implies that Yang's space-time algebra (YSTA) presupposes for its representation space 
to take representation bases like 
\begin{eqnarray}
| t/\lambda,n_{12}, \cdots> \equiv |{\hat{\Sigma}}_{0a} =t/\lambda> |{\hat{\Sigma}}_{12}=n_{12}>
\cdots|{\hat{\Sigma}}_{910}=n_{910}>,
\end{eqnarray}
where $t$ denotes {\it time}, the continuous eigenvalue of $\hat{X}_0 \equiv \lambda\ \hat{\Sigma}_{0 a}$ 
and $n_{12}, \cdots$ discrete eigenvalues of maximal commuting set of subalgebra of $SO(D+1,1)$ which are 
commutative with ${\hat{\Sigma}}_{0a}$, for instance, ${\hat{\Sigma}}_{12}$, ${\hat{\Sigma}}_{34},\cdots , 
{\hat{\Sigma}}_{910}$, when $D=11$.$^{[9],[1],[2]}$

Indeed, an infinite dimensional linear space expanded by $|\ t/\lambda, 
n_{12},\cdots>$ mentioned above provides a representation space of unitary infinite dimensional representation of YSTA. It is 
the so-called "quasi-regular representation"$^{[10]}$ of SO(D+1,1),\footnote{It corresponds, in the case of unitary 
representation of Lorentz group $SO(3,1)$, to taking $K_3\ (\sim \Sigma_{03})$ and $J_3\ (\sim \Sigma_{12})$ to be 
diagonal, which have continuous and discrete eigenvalues, respectively, instead of ${\bf J}^2$ and $J_3$ in 
the familiar representation.}
and is decomposed into the infinite series of the ordinary unitary irreducible representations of 
$SO(D+1,1)$ constructed on its maximal compact subalgebra, $SO(D+1)$. 

It means that there holds the following form of decomposition theorem,
\begin{eqnarray}
| t/\lambda, n_{12},\cdots>= \sum_{\sigma 's}\ \sum_{l,m}\  C^{\sigma's, n_{12}, \cdots }_{l,m}(t/\lambda)\ 
 | \sigma 's ; l,m>,
\end{eqnarray}      
with expansion coefficients $C^{\sigma's, n_{12}, \cdots}_{l,m}(t/\lambda).^{[9],[1]}$ In Eq.(2.13), 
$|\sigma 's ; l, m>'s$ on the right hand side describe the familiar unitary irreducible representation 
bases of $SO(D+1,1)$, which are designated by $\sigma 's$ and $(l,m),$  
\footnote{In the familiar unitary irreducible representation of $SO(3,1)$, it is well known that $\sigma$'s are 
represented by two parameters, $(j_0, \kappa)$, with $j_0$ being $1,2, \cdots \infty$ and $\kappa$ being purely 
imaginary number, for the so-called principal series of representation. With respect to the associated representation 
of $SO(3)$, when it is realized on $S^2$, as in the present case, $l$'s denote positive integers, 
$l= j_0, j_0+1, j_0+2,\cdots,\infty$, and $m$ ranges over $\pm l, \pm(l-1) , \cdots,\pm1, 0.$ } 
denoting, respectively, the irreducible unitary representations of $SO(D+1,1)$ and the associated 
irreducible representation bases of $SO(D+1)$, the maximal compact subalgebra of $SO(D+1,1)$, mentioned above. 
It should be noted here that, as remarked in I, $l$'s  are limited to be integer, excluding the possibility of 
half-integer, because of the fact that generators of $SO(D+1)$ in YSTA are defined as differential operators 
on $S^D$, i.e., ${q_1}^2 + {q_2}^2 + \cdots + {q_{D-1}}^2 + {q_a}^2 + {q_b}^2 = 1.$

In what follows, let us call the infinite dimensional representation space introduced above for the representation of 
YSTA, Hilbert space I, in distinction to Hilbert space II which is Fock-space constructed dynamically by 
creation-annihilaltion operators of second-quantized fields on YSTA, such as $D_0$ brane field,$^{[9]}$ discussed 
in section 4.

\section{\normalsize Kinematical Holographic Relation [KHR] in YSTA}

\subsection{\normalsize Recapitulation of Kinematical Holographic Relation [KHR]} 

First, let us remember that the following kinematical holographic relation\footnote{The argument in this 
subsection was given in I, ref. [1], on the basis of the following form of $D_0$ brane field equation: 
$[({X_\sigma}^2 +R^2 N^2 )( (\partial/\partial {X_\mu})^2 + R^{-2} (\partial/\partial {N})^2)) 
 - ( X_\mu \partial/\partial {X_\mu} + N \partial/\partial{N})^2 - (D-1)( X_\mu \partial/\partial {X_\mu}+N
 \partial/\partial{N})\ ]\ D ( X_\nu, N) = 0,$ which was derived in ref. [9] from the following $D_0$ brane field 
action after M-theory,$^{[7]}$ 
$ \bar{\hat L} = A\ {\rm tr}\  \{ [\hat {\Sigma}_{KL}, 
\hat {D}^\dagger]\ [\hat {\Sigma}_{KL}, \hat {D}]\}=  A'\ {\rm tr}\  \{ 2\ (R^2 /\hbar^2)\  [{\hat P}_\mu, 
\hat {D}^\dagger ]\ [ \hat {P}_\mu, \hat {D}] - {\lambda}^{-4 }\ [\ [\hat {X}_\mu, \hat {X}_\nu], \hat {D}^\dagger] 
[ [\hat {X}_\mu,\hat {X}_\nu], \hat {D}]\},$ with $K, L = (\mu, b),$ by means of the Moyal star product method.} 
\begin{eqnarray}
\hspace{-3cm} [KHR] \hspace{2cm}       n^L_{\rm dof}= {\cal A} / G,
\end{eqnarray}
with the proportional constant $G$
\begin{eqnarray}
G\  \sim {(2 \pi)^{D/2} \over 2}\ (D-1)!! &&for\ D\ even
\\
    \sim  (2 \pi)^{(D-1)/2}(D-1)!! &&for\ D\ odd,
\end{eqnarray}
was derived in I for the $D$-dimensional space-like region with finite radius $L$ in D-dimensional 
Yang's quantized space-time in the unit of $\lambda$. Let us denote the region hereafter as $V_D^L$, 
which was defined by

\begin{eqnarray}
 \sum_{K \neq 0}{\Sigma_{aK}}^2  = \sum_{\mu \neq 0}{\Sigma_{a \mu}}^2 + {\Sigma_{ab}}^2 = (L/\lambda)^2,
\end{eqnarray}
or
\begin{eqnarray}
{X_1}^2 + {X_2}^2 + \cdots + {X_{D-1}}^2 + R^2\ N^2 = L^2.
\end{eqnarray}
Here, $\Sigma_{MN}$'s are presumed to be given in terms of Moyal star product formalism applied to the 
expression, $\Sigma_{MN}= ( -q_M p_n + q_N p_M)$, as was  treated in detail in I.

${\cal A}$ in\ [KHR] (3.1) simply denotes the boundary surface area of $V_D^L$, that is,  
\begin{eqnarray}
{\cal A} =  ({\rm area\ of}\ S^{D-1}) ={(2 \pi)^{D/2} \over {(D-2)!!}} (L/\lambda)^{D-1} 
&&for\ D\ even,
\nonumber\\                   =2 {(2\pi)^{(D-1)/2} \over {(D-2)!!}} (L/\lambda)^{D-1} &&for\ D\ odd. 
\end{eqnarray}

On the other hand, $n^L_{\rm dof}$ in [KHR] (3.1), which denotes, by definition, the number of spatial 
degrees of freedom of YSTA inside $V_D^L$, was given in I as follows,  
\begin{eqnarray}
&&n^L_{\rm dof}  =  dim\ ( \rho_{[L/\lambda]}) = {2 \over (D-1)!}{([L/\lambda]+ D-2)! \over ([L/\lambda]-1)!}
\nonumber\\
 &&\hskip3.5cm \sim {2 \over (D-1)!}  [L/\lambda]^{D-1}.
\end{eqnarray}

Indeed, the derivation of the above equation (3.7) was the central task in I. In fact, we emphasized that 
the number of degrees of freedom $n^L_{\rm dof}$ inside $V_D^L$, which is subject to noncommutative algebra, YSTA, 
should be, logically and also practically, found in the structure of representation space of YSTA, that is, 
Hilbert space I defined in section 2. Let us here recapitulate in detail the essence of the derivation in order to 
make the present paper as self-contained as possible. 
   
In fact, one finds that the representation space needed to calculate $n^L_{\rm dof}$ is 
prepared in Eq.(2.13), where any "quasi-regular" representation basis,\\ $ | t/\lambda, n_{12}, \cdots>$, 
is decomposed into the infinite series of the ordinary unitary representation bases of 
$SO(D+1,1)$, $| \sigma 's ; l,m>.$ 
As was stated in subsection 2.2, the latter representation bases, $| \sigma 's ; l,m>'s$ are constructed on 
the familiar finite dimensional representations of maximal compact subalgebra of YSTA, $SO(D+1)$, whose representation 
bases are labeled by $(l,m)$ and provide the representation bases for spatial quantities under consideration, 
because $SO(D+1)$ just involves those spatial operators $( \hat{X}_u, R \hat{N})$. 

In order to arrive at the final goal of counting $n^L_{\rm dof}$, therefore, one has only to find mathematically 
a certain irreducible representation of $SO(D+1)$, which {\it properly} describes (as seen in what follows) 
the spatial quantities $( \hat{X}_u, R \hat{N})$ inside the bounded region with radius $L$, then one finds 
$n^L_{\rm dof}$ through counting the dimension of the representation. 

At this point, it is important to note that, as was remarked in advance in subsection 2.2, any generators of $SO(D+1)$ 
in YSTA  are defined by the differential operators on the $D-$dimensional unit sphere, $S^D$, i.e., 
${q_1}^2 + {q_2}^2 + \cdots + {q_{D-1}}^2 + {q_a}^2 + {q_b}^2 = 1,$ limiting its representations with $l$ to be 
integer.
   
On the other hand, it is well known that the irreducible representation of arbitrary high-dimensional $SO(D+1)$ 
on $S^D = SO(D+1)/SO(D)$ is derived in the algebraic way, $^{[11]}$ irrelevantly to any detailed knowledge of 
the decomposition equation (2.13), but solely in accord with the fact that $SO(D+1)$ in YSTA is defined originally 
on $S^D$, as mentioned above. One can choose, for instance, $SO(D)$ with generators $\hat{\Sigma}_{MN} (M,N=b, u)$, 
while $SO(D+1)$ with generators $\hat{\Sigma}_{MN}(M,N=a,b,u)$. Then, it turns out that any irreducible representation 
of $SO(D+1)$, denoted by $\rho_l$, is uniquely designated by the maximal integer $l$ of eigenvalues of 
${\hat \Sigma}_{ab}$ in the representation, where ${\hat \Sigma}_{ab}$ is known to be a possible Cartan subalgebra 
of the so-called compact symmetric pair $(SO(D+1),SO(D))$ of rank $1$.$^{[11]}$ 

According to the so-called Weyl's dimension formula, the dimension of $\rho_l$ is given by$^{[11],[1],[2]}$
\begin{eqnarray}
 dim\ (\rho_l)= {(l+\nu) \over \nu} {(l+2\nu-1)! \over {l!(2\nu -1)!}},
\end{eqnarray}
where $ \nu \equiv (D-1)/2$ and $D \geq 2$.\footnote{This equation just gives the familiar result $dim\ (\rho_l)= 2l +1,$ 
in the case $SO(3)$ taking $D=2.$} 

Finally, we can find a certain irreducible representation of $SO(D+1)$ among those $\rho_l 's $  given above, 
which {\it properly} describes (or realizes) the spatial quantities inside the bounded region $V_D^L$. 
Now, let us choose tentatively $l = [L/\lambda]$ with $[L/\lambda]$ being the integer part of $L/\lambda$. In this case, 
one finds out that the representation $\rho_{[L/\lambda]}$ just {\it properly} describes all of generators of 
$SO(D+1)$ inside the above bounded spatial region $V_D^L$, because  $[L/\lambda ]$ indicates also the largest eigenvalue 
of any generators of $SO(D+1)$ in the representation $\rho_{[L/\lambda]}$ on account of its $SO(D+1)-$invariance and 
hence eigenvalues of spatial quantities $( \hat{X}_u, R \hat{N})$ are well confined inside the bounded region with 
radius {L}. As the result, one finds that the dimension of $\rho_{[L/\lambda]}$ just gives the number of spatial degrees 
of freedom inside $V_D^L$, $n^L_{\rm dof}$, as shown in (3.7).

\subsection{\normalsize KHR in the lower-dimensional spatial region $V_d^L$ }
According to the argument given for $V_D^L$ in the preceding subsection, let us study the 
kinematical holographic relation in the lower-dimensional bounded spatial region $V_d^L$ for the subsequent 
argument of the area-entropy relation in section 4. In fact, it will be given through a simple $D_0$ brane gas system 
formed inside $d\ (\leq{D-1})$-dimensional bounded spatial region, $V_d^L$, which is defined 
by
\begin{eqnarray}  
{X_1}^2 + {X_2}^2 + \cdots + {X_d}^2 = L^2,
\end{eqnarray}
instead of (3.5).

In this case, the boundary area of $V_d^L$, that is, ${\cal A}\ (V_d^L)$ is given by
\begin{eqnarray}
{\cal A}\ (V_d^L) =  ({\rm area\ of}\ S^{d-1}) ={(2 \pi)^{d/2} \over {(d-2)!!}} (L/\lambda)^{d-1} 
&&for\ d\ even
\nonumber\\                   =2 {(2\pi)^{(d-1)/2} \over {(d-2)!!}} (L/\lambda)^{d-1} &&for\ d\ odd, 
\end{eqnarray}
corresponding to Eq. (3.6).

On the other hand, the number of degrees of freedom of $V_d^L$, let us denote it $n_{dof} (V_d^L)$, is calculated 
by applying the arguments given for derivation of $n_{dof}^L$ in (3.7). In fact, it is found in a certain irreducible 
representation of $SO(d+1)$, a minimum subalgebra of YSTA, which includes the $d$ spatial quantities, $\hat{X}_1, 
\hat{X}_2, \cdots, \hat{X}_d$ needed to properly describe $V_d^L$, and is really constructed by the generators 
$\hat{\Sigma}_{MN}$ with $M,N$ ranging over $a,1,2, \cdots,d$. The representation of $SO(d+1)$, let us denote it 
$\rho_l\ (V_d^L)$ with suitable integer $l = [L/\lambda]$, is given on the representation space $S^d =SO(d+1)/SO(d)$, 
taking the subalgebra $SO(d)$, for instance, $\hat{\Sigma}_{MN}$ with $M,N$ ranging over 
$1,2,\cdots,d$, entirely in accord with the argument on the irreducible representation of $SO(D+1)$ given in the 
preceding subsection 3.1. 

One immediately finds that 
\begin{eqnarray}
&&n_{\rm dof}\ (V_d^L)  =  dim\ ( \rho_{[L/\lambda]}\ (V_d^L)) = {2 \over (d-1)!}{([L/\lambda]+ d-2)! \over ([L/\lambda]-1)!}
\nonumber\\
 &&\hskip3.5cm \sim {2 \over (d-1)!}  [L/\lambda]^{d-1}.
\end{eqnarray}
corresponding to (3.7), and there holds, from (3.10) and (3.11), the following kinematical holographic 
relation for $V_d^L$ in general
\begin{eqnarray}
\hspace{-3cm} [KHR] \hspace{2cm}       n^L_{\rm dof}\ (V_d^L)=
{\cal A}\ (V_d^L) / G_d,
\end{eqnarray}
with the proportional constant $G_d$
\begin{eqnarray}
G_d \sim {(2 \pi)^{d/2} \over 2}\ (d-1)!! &&for\ d\ even
\\
    \sim  (2 \pi)^{(d-1)/2}(d-1)!! &&for\ d\ odd,
\end{eqnarray}
corresponding to Eqs. (3.1)- (3.3) for $V_D^L$.  
     
\section{\normalsize Area-Entropy Relation in $D_0$ Brane Gas subject to YSTA}
  
\subsection{\normalsize $D_0$ Brane Gas Model in $V_d^L$ and Its Mass and Entropy}
Now, let us consider the central problem of the present paper, that is, the derivation of a possible area-entropy relation 
through a simple $D_0$ brane gas$^{[6]}$ model formed inside $V_d^L$ according to the idea of M-theory. 
This implies that one has to deal with the dynamical system of the second-quantized $D_0$ brane field ${\hat D}_0$ 
inside $V_d^L$. In the present toy model of the $D_0$ brane gas, however, we avoid to enter into detail of the dynamics 
of $D_0$ brane system, but treat it as an ideal gas, only taking into consideration 
that the system is developed on $V_d^L$ subject to YSTA and its representation discussed above, but neglecting 
interactions of $D_0$ branes, for instance, with strings, as well as possible self-interactions among themselves. 

First of all, according to the argument given in the preceding subsection 3.2, the spatial structure of $V_d^L$ 
is described through the specific representation $ \rho_{[L/\lambda]}\ (V_d^L)$. Let us denote its orthogonal 
basis-vector system in Hilbert space I, as follows 
\begin{eqnarray}
 \rho_{[L/\lambda]}\ (V_d^L): \quad |\ m >,  \qquad  m= 1,2,\cdots, n_{\rm dof}(V_d^L).     
\end{eqnarray}   
 In the above expression, $n_{\rm dof}\ (V_d^L)$ denotes the dimension of the representation 
$\rho_{[L/\lambda]}\ (V_d^L)$, as defined in (3.11).

At this point, one should notice that the {\it second quantized} ${\hat D}_0$-brane field$^{[12]}$ on $V_d^L$ 
must be the linear operators operating on Hilbert space I, and described by $n_{\rm dof}(V_d^L) \times n_{\rm dof}(V_d^L)$ 
matrix under the representation $\rho_{[L/\lambda]}\ (V_d^L)$ like $< m\ |{\hat D}|\ n >$ on the one hand, and 
on the other hand each matrix element must be operators operating on Hilbert space II, playing the role of 
creation-annihilation of $D_0$ branes. On the analogy of the ordinary quantized local field, let us define 
those creation-annihilation operators through the diagonal parts in the following way:\footnote{On the other hand, 
the non-diagonal parts, $< m\ |{\hat D}|\ n >,$ 
are to be described in terms like ${\bf a}_m {\bf a}_n^\dagger$ or ${\bf a}_m^\dagger {\bf a}_n$ in accord with 
the idea of M-theory where they are conjectured to be concerned with the interactions between $[site\ m]$ and $[site\ n].$ 
The details must be left to the rigorous study of the second quantization of $D_0$-brane field.$^{[12]}$}    
\begin{eqnarray}
< m\ | {\hat D}|\ m >\  \sim\ {\bf a}_m\ {\rm or}\  {\bf a}_m^\dagger.
\end{eqnarray}
In the above expression, ${\bf a}_m$ and ${\bf a}_m^\dagger$, respectively, denote annihilation and creation 
operators of $D_0$ brane, satisfying the familiar commutation relations,
\begin{eqnarray}
&&[{\bf a}_m, {\bf a}_n^\dagger]= \delta_{mn},
\\
&&[{\bf a}_m, {\bf a}_n]= 0 .
\end{eqnarray}
 One notices that the labeling number $m$ of basis vectors, which ranges from $1$ to $n_{\rm dof}(V_d^L)$ 
plays the role of {\it spatial coordinates} of $V_d^L$ in the present noncommutative YSTA, corresponding to the so-called 
lattice point in the lattice theory. Let us denote the {\it point} hereafter $[site]$ or $[site\ m]$ of $V_d^L$.

Now, let us focus our attention on quantum states constructed dynamically in Hilbert space II by the creation-annihilation 
operators ${\bf a}_m$ and ${\bf a}_m^\dagger$ of $D_0$ branes introduced above at each [site] inside $V_d^L$. One should 
notice here the important fact that in the present simple $D_0$ brane gas model neglecting all interactions of $D_0$ branes, 
each $[site]$ can be regarded as independent quantum system and described in general by own statistical operator, while 
the total system of gas is described by their direct product. In fact, the statistical operator at each $[site\  m]$ 
denoted by ${\hat W}[m]$, is given in the following form,
\begin{eqnarray} 
{\hat W}[m] = \sum_ k  w_k\  |\ [m]: k >\  < k :[m]\ |,
\end{eqnarray}
with 
\begin{eqnarray}
 |\ [m]:  k  >  \equiv {1 \over \sqrt{k!}}({\bf a}_m^\dagger)^{k}|\ [m]:0 >.
\end{eqnarray}
That is, $|\ [m]:  k > ( k=0, 1, \cdots)$ describes the normalized quantum-mechanical state in Hilbert space II with 
$k$ $D_0$ branes constructed by ${\bf a}_m^\dagger$ on $|\ [m]:0 >,$ i.e. the vacuum state of $[site\ m]$.
\footnote{The proper vacuum state in Hilbert space II is to be expressed by their direct product.} And $w_k$'s denote 
the realization probability of state with occupation number $k$, satisfying $\sum_k w_k = 1.$ 

We assume here that the statistical operator at each $[site\ m]$ is common to every [site] in the present $D_0$ 
brane gas under equilibrium state, with the common values of $w_k$'s and the statistical operator of total 
system on $V_d^L$ , ${\hat W}(V_d^L)$, is given by  
\begin{eqnarray}
{\hat W}(V_d^L) = {\hat W}[1] \otimes {\hat W}[2] \cdots \otimes {\hat W}[m] \cdots \otimes {\hat W}[n_{dof}].
\end{eqnarray}

Consequently, one finds that the entropy of the total system, $S(V_d^L)$, is given by
\begin{eqnarray}
S(V_d^L) = - {\rm Tr}\ [{\hat W}(V_d^L)\ {\rm ln} {\hat W}(V_d^L)] = n_{dof}(V_d^L)\times S[site], 
\end{eqnarray}
where $S[site]$ denotes the entropy of each [site] assumed here to be common to every [site] and given by
\begin{eqnarray}
S[site] = - {\rm Tr}\ [ {\hat W}[site]\ {\rm ln}{\hat W}[site]] = - \sum_k w_k\  {\rm ln} w_k.
\end{eqnarray}

Comparing this result (4.8) with [KHR]\ (3.12) derived in the preceding section, we find an important fact that the 
entropy $S(V_d^L)$ is proportional to the surface area ${\cal A}\ (V_d^L)$, that is, a kind of area-entropy relation 
([AER]) of the present system:
\begin{eqnarray}
\hspace{-3cm} [AER] \hspace{2cm}         S(V_d^L) = {\cal A}\ (V_d^L)\ {S[site] \over G_d},
\end{eqnarray}
where $G_d$ is given by (3.13)-(3.14). 

Next, let us introduce the total energy or mass of the system, $M(V_d^L)$. If one denotes the average energy or mass 
of the individual $D_0$ brane inside $V_d^L$ by $\mu$, it may be given by
\begin{eqnarray}   
M(V_d^L) = \mu {\bar N}[site]\  n_{\rm dof}(V_d^L) \sim  \mu {\bar N}[site] {2  \over (d-1)!}  [L/\lambda]^{d-1},
\end{eqnarray}
where ${\bar N}[site]$ denotes the average occupation number of $D_0$ brane at each $[site]$ given by
\begin{eqnarray}
{\bar N}[site] \equiv \sum_k k w_k.
\end{eqnarray}

Comparing this expression (4.11) with (4.8) and (3.12), respectively, we obtain a kind of mass-entropy relation ([MER]) 
\begin{eqnarray}
\hspace{-2cm} [MER] \hspace{2cm}         M(V_d^L) / S(V_d^L) = \mu {\bar N}[site] / S[site],
\end{eqnarray}
and  a kind of area-mass relation ([AMR])
\begin{eqnarray}
\hspace{-3cm} [AMR] \hspace{2cm} M(V_d^L) = {\cal A}(V_d^L)\ {\mu {\bar N}[site] \over G_d}.
\end{eqnarray}
     
\subsection{\normalsize Schwarzschild Black Hole and Area-Entropy Relation In $D_0$ brane Gas System}

In the preceding subsection 4.1, we have studied $D_0$ brane gas system and derived area-entropy relation $[AER]$ 
(4.10), mass-entropy and area-mass relations, $[MER]$ (4.13) and $[AMR]$ (4.14), which are essentially based on 
the kinematical holographic relation in YSTA studied in section 3. 

At this point, it is quite important to notice that these three relations explicitly depend on 
the following ${\it static}$ factors of the gas system, $\mu$, ${\bar N}[site]$ and $S[site]$, that is, the average 
energy of individual $D_0$ brane, the average occupation number of $D_0$ branes and the entropy at each [site], 
which are assumed to be common to every [site], while these factors turn out to play an important role in 
arriving finally at the area-entropy relation in connection with black holes, as will be seen below.   

Now, let us investigate how the present gas system tends to a black hole. We assume for simplicity that the system 
is under $d=3$, and becomes a Schwarzschild black hole, in which the above factors acquire certain limiting 
values, $ \mu_S$, ${\bar N}_S[site]$ and $S_S [site]$, while the size of the system, $L$, becomes $R_S$, 
that is, the so-called Schwarzschild radius given by 
\begin{eqnarray}   
R_S \equiv  2 G M(V_3^{R_S})/c^2,
\end{eqnarray}
where $G$ and $c$ denote Newton 's constant and the light velocity, respectively, and $M(V_3^{R_S})$ is given by 
Eq.(4.11) with $L= R_S$, $\mu = \mu_S$ and ${\bar N}[site] = {\bar N}_S[site]$. Indeed, inserting the 
above values into Eq.(4.11), we arrive at the important relation, called hereafter the black hole condition [BHC], 
\begin{eqnarray}
\hspace{-1cm} [BHC] \hspace{2cm} M(V_3^{R_S}) = {\lambda^2 \over 4\mu_S {\bar N}_S [site]}{c^4 \over G^2} 
= {M_P^2 \over 4 \mu_S {\bar N}_S[site]}.
\end{eqnarray}
In the last expression, we assumed that $\lambda$, i.e., the small scale parameter in YSTA is equal  
to Planck length $l_P = [G \hbar / c^3]^{1/2} = \hbar /( c M_P )$, where $M_P$  denotes Planck mass.

On the other hand, we simply obtain the area-entropy relation [AER] under the Schwarzschild black hole by 
inserting the above limiting values into [AER] (4.10)   
\begin{eqnarray}
S(V_3^{R_S}) = {\cal A}\ (V_3^{R_S})\ {S_S[site] \over 4 \pi},
\end{eqnarray}
noting that $G_3 =4 \pi.$

At this point, one finds that it is a very important problem how to relate the above area-entropy relation 
under a Schwarzschild black hole with [AER] (4.10) of $D_0$ brane gas system in general, which is 
derived irrelevantly of the detail whether the system is a black hole or not. As was mentioned in the beginning 
of this subsection, however, the problem seems to exceed the applicability limit of the present toy model 
of $D_0$ brane gas, where the system is treated solely as a {\it static} state under {\it given} values of 
parameters, $\mu$, ${\bar N}[site]$ and $S[site]$, while the critical behavior around the formation of Schwarzschild 
black hole must be hidden in a possible ${\it dynamical}$ change of their values. 

In order to supplement such a defect of the present static toy model, let us try here a Gedanken-experiment, 
in which one increases the entropy of the gas system $S(V_3^L)$, keeping its size $L$ at the initial value $L_0$, 
until the system tends to a Schwarzschild black hole, where Eqs.(4.16) and (4.17) with $R_S = L_0$ hold. Then, one 
finds that according to [AER] (4.10), the entropy of $[site]$, $S[site]$ increases proportionally to $S(V_3^L)$ and 
reaches the limiting value $S_S [site]$, starting from any initial value $S_0[site]$ prior to formation of the black 
hole, because ${\cal A}(V_d^L)$ in Eq.(4.10) is invariant during the process. Namely, one finds a very simple fact 
that $S_0[site] \leq S_S [site].$  However, this simple fact combined with [AER] (4.10) leads us to the following 
form of a new area-entropy relation which holds throughout for the $D_0$ brane gas system up to the formation of 
Schwarzschild black hole,\footnote{Similarly, by the second Gedanken-experiment, in which one increases 
the total mass of gas system $M(V_3^L)$ with the fixed size $L_0$ in connection with [AMR] (4.14), 
in place of the increase of the entropy of gas system $S(V_3^L)$ in the first Gedanken-experiment, 
one obtains a new area-mass relation [AMR], $M(V_3^L) \leq {\cal A}(V_3^L) \mu_S {\bar N}_S[site] / 4 \pi$.}
\begin{eqnarray}
\hspace{-3cm} [AER] \hspace{2cm}  S(V_3^L) \leq {{\cal A} (V_3^L) S_S[site] \over 4 \pi},
\end{eqnarray}
where the equality holds for Schwarzschild black hole, as seen in Eq. (4.17).

\section{\normalsize Concluding Remarks}

We have derived the area-entropy relation [AER] (4.18) together with (4.10) in our toy model of $D_0$ brane gas subject to Yang's quantized 
space-time algebra, YSTA. Indeed, it is essentially based on the fundamental nature of the noncommutative geometry
of YSTA, that is, the kinematical reduction of spatial degrees of freedom and holographic relation in YSTA, 
which was pointed out in I and now extended to the lower dimensional region, $V_d^L$, as shown by [KHR] (3.12) 
in section 3. In addition, it should be noted that [AER] (4.18) has been derived with aid of a crude Gedanken-experiment 
on $D_0$ brane gas system. 

Before entering into discussions on the implication of the relation, let us consider [AER] (4.18) in comparison with the 
Bekenstein-Hawking area-entropy relation or holographic principle, which has been discussed in various ways 
during the past decades and typically shown in the simple form as $S \leq {\cal A} /4$ in accord with 
the expression in [AER] (4.18). First of all, one finds out that the latter relation provides an important 
knowledge on the present [AER] (4.18), 
that is,
\begin{eqnarray}       
 S_S[site] /4 \pi = 1/4
\end{eqnarray}
or
\begin{eqnarray}       
 S_S[site] = \pi .
\end{eqnarray}
If one remembers that $S[site]$ can be expressed in general by
\begin{eqnarray}
S[site] = {\rm ln}\ {\cal N}[site]
\end{eqnarray}
with ${\cal N}[site]$ the number of independent quantum states in $[site]$ or, roughly speaking, the maximal limit 
of occupation number $k$ in Eqs. (4.5)-(4.6), one finds out that Eq. (5.2) implies
\begin{eqnarray}
{\cal N}_S [site] = e^\pi.
\end{eqnarray}
 
At this point, there arises a fundamental question why our present approach is unable to derive such a condition 
(5.2) or (5.4). It is certainly clear, because of the applicability of the present toy model of $D_0$ brane gas 
where the system is treated as a kind of ideal gas of $D_0$ branes neglecting all of their interactions, as was 
stated in the beginning of section 4. It is well-known that the factor {1/4} on the right hand side of (5.1) comes 
from the consideration of Hawking radiation of black holes. 

Furthermore, the present approach does not answer the question what happens when $S[site]$ exceeds $S_S [site]$. 
It is surely concerned with the problem of gravitational collapse or evaporation of black hole. In order 
to answer these questions, therefore, it is clear that one has to take into consideration at least the interactions 
of $D_0$ branes with gravitation- or radiation- fields, which are expected to come from interactions of $D_0$ branes 
with open and closed strings. All of them, however, exceed the scope of the present paper. We have tried here solely 
to give an outline to arrive at area-entropy relation in our present scheme, leaving the satisfactory treatment of the 
quantum field theory of $D_0$ brane$^{[12]}$ subject to YSTA to the forthcoming paper, in which we expect 
further that the theory is ultimately free from UV- and IR-divergences.$^{[1],[2],[13]}$

\end{document}